\documentclass[tighten,preprint2,flushrt]{aastex}
\usepackage{geometry}
\usepackage{float}
\usepackage{slashbox}
\usepackage{multirow}
\usepackage[hyphens]{url}
\usepackage{subfigure}
\usepackage{color}
\usepackage{color,soul}
\usepackage{rotating}
\usepackage{enumerate}
\usepackage{caption2}
\usepackage{amsmath}
\usepackage{autobreak}
\newcommand{\myemail}{zouhu@nao.cas.cn}

\begin{document}
\title{The Second Data Release of the Beijing--Arizona Sky Survey}
%\correspondingauthor{Hu Zou}
%\email{zouhu@nao.cas.cn, zouhu0628@gmail.com}

\author{Hu Zou\altaffilmark{1}, Tianmeng Zhang\altaffilmark{1,2}, Zhimin Zhou\altaffilmark{1}, Xiyan Peng\altaffilmark{1}, Jundan Nie\altaffilmark{1}, Xu Zhou\altaffilmark{1}, Xiaohui Fan\altaffilmark{3}, Linhua Jiang\altaffilmark{4}, Ian McGreer\altaffilmark{3}, Arjun Dey\altaffilmark{5}, Dongwei Fan\altaffilmark{1}, Joseph R. Findlay\altaffilmark{6}, Jinhua Gao\altaffilmark{1}, Yizhou Gu\altaffilmark{7}, Yucheng Guo\altaffilmark{4,8}, Boliang He\altaffilmark{1}, Junjie Jin\altaffilmark{1}, Xu Kong\altaffilmark{9}, Dustin Lang\altaffilmark{10}, Fengjie Lei\altaffilmark{1}, Michael Lesser\altaffilmark{3}, Feng Li\altaffilmark{7}, Jun Ma\altaffilmark{1,2}, Xiaolei Meng\altaffilmark{11}, Moe Maxwell\altaffilmark{12}, Adam D. Myers\altaffilmark{6}, Liming Rui\altaffilmark{11}, David Schlegel\altaffilmark{13}, Fengwu Sun\altaffilmark{4,8}, Hong Wu\altaffilmark{1}, Jiali Wang\altaffilmark{1}, Qirong Yuan\altaffilmark{7}} 

\altaffiltext{1}{Key Laboratory of Optical Astronomy, National Astronomical Observatories, Chinese Academy of Sciences, Beijing 100012, People's Republic of China; \myemail}
\altaffiltext{2}{College of Astronomy and Space Sciences, University of Chinese Academy of Sciences, Beijing 100049, People's Republic of China}
\altaffiltext{3}{Steward Observatory, University of Arizona, Tucson, AZ 85721, USA}
\altaffiltext{4}{Kavli Institute for Astronomy and Astrophysics, Peking University, Beijing 100871, People's Republic of China}
\altaffiltext{5}{National Optical Astronomy Observatory, Tucson, AZ 85719, USA}
\altaffiltext{6}{Department of Physics and Astronomy, University of Wyoming, Laramie, WY 82071, USA}
\altaffiltext{7}{Department of Physics, Nanjing Normal University, WenYuan Road 1, Nanjing 210046, People's Republic of China}
\altaffiltext{8}{Department of Astronomy, School of Physics, Peking University, Beijing 100871, People's Republic of China}
\altaffiltext{9}{Key Laboratory for Research in Galaxies and Cosmology, Department of Astronomy, University of Science and Technology of China, Hefei 230026, People's Republic of China}
\altaffiltext{10}{David Dunlap Institute, University of Toronto, Toronto, Canada}
\altaffiltext{11}{Department of Physics and Tsinghua Center for Astrophysics, Tsinghua University, Beijing 100086, People's Republic of China}
\altaffiltext{12}{Harvard-Smithsonian Center for Astrophysics, 60 Garden Street, Cambridge, MA 02138, USA}
\altaffiltext{13}{Lawrence Berkeley National Labortatory, Berkeley, CA 94720, USA}

%%%%%%%%%%%%%%%%%%%%%%%%
\begin{abstract} 
This paper describes the second data release (DR2) of the Beijing--Arizona Sky Survey (BASS).  BASS is an imaging survey covering a 5400 deg$^2$ footprint in the $g$ and $r$ bands using the 2.3 m Bok telescope. DR2 includes the observations through 2017 July obtained by BASS and by the Mayall $z$-band Legacy Survey (MzLS), which used the 4 m Mayall telescope to observe the same footprint. BASS and MzLS have completed 72\% and 76\% of their observations. The two surveys will be served for the spectroscopic targeting of the upcoming Dark Energy Spectroscopic Instrument. Both BASS and MzLS data are reduced by the same pipeline. We have updated the basic data reduction and photometric calibrations in DR2. In particular, source detections are performed on stacked images, and photometric measurements are co-added from single-epoch images based on these sources. The median 5$\sigma$ point-source depths after Galactic extinction corrections are 24.05, 23.61, and 23.10 mag for the $g$, $r$, and $z$ bands, respectively. The DR2 data products include stacked images, co-added catalogs, and single-epoch images and catalogs. The BASS website (\url{http://batc.bao.ac.cn/BASS/}) provides detailed information and links to download the data.

\end{abstract}

\keywords{surveys --- techniques: image processing --- techniques: photometric}

%%%%%%%%%%%%%%%%%%%%%%%%%
\section{Introduction}
Large-scale spectroscopic surveys over the past 20 years have revolutionized our view of the universe from the structure of the Milk Way and galaxy evolution to large-scale structure and dark energy \citep{col01,yor00,sco07,san12,zha12,new13,lis15}. The Sloan Digital Sky Survey \citep[SDSS;][]{yor00} began with spectroscopy on the brightest targets in its imaging data.  SDSS has since performed spectroscopy on progressively deeper targets, with SDSS-III/BOSS and SDSS-IV/eBOSS pushing to the faint limits of that imaging for the measurement of baryonic acoustic oscillations \citep{daw13,daw16}. Deeper wide-field imaging data is required prior to the era of the Large Synoptic Survey Telescope \citep[LSST;][]{lss09} to select targets for upcoming wide-field spectroscopic surveys, such as the Dark Energy Spectroscopic Instrument \citep[DESI;][]{des16} and Subaru Prime Focus Spectrograph \citep[PFS;][]{tak14}.

DESI is a next-generation dark energy experiment that will accurately measure the expansion rate and structure growth of the universe. It will obtain spectroscopic redshifts of about 34 million galaxies and quasars, which is one magnitude more than those of the SDSS \citep{des16}. In addition to cosmology, the DESI project will play an important role in shaping our understanding of the Milky Way and galaxy evolution by observing stars and low-redshift galaxies during bright time. 

The spectroscopic target selection for DESI will be based on three optical bands ($g$, $r$, and $z$) and two near-infrared bands ($W1$ and $W2$). The optical imaging surveys include three components: the Beijing--Arizona Sky Survey \citep[BASS;][]{zou17b}, the Dark Energy Camera Legacy Survey \citep[DECaLS;][]{blu16}, and the Mayall $z$-band Legacy Survey \citep[MzLS;][]{sil16}. DECaLS covers a 9000 deg$^2$ equatorial footprint using the Dark Energy Camera (DECam) on the 4 m Blanco telescope at CTIO. BASS and MzLS, using the Bok and Mayall Telescopes, respectively,  on Kitt Peak, are covering an adjacent 5000 deg$^2$ footprint in the north Galactic cap. All three optical surveys share a similar observing strategy. Dynamic exposure time calculators (ETCs) ensure their imaging depths will satisfy the requirement of the DESI spectroscopic target selection. The near-infrared images at wavelengths of 3.6 $\mu$m ($W1$) and 4.5 $\mu$m ($W2$) make use of the full-sky data from the $Wide$-$field$ $Infrared$ $Survey$ $Explorer$ mission \citep[WISE;][]{wri10} and the $NEOWISE$-$Reactivation$ data \citep{mai14}, coadded to their full depth \citep{mei17,mei18}.
%The near-infrared $W1$ and $W2$ data come from the latest survey of the Wide-field Infrared Survey Explorer \citep[WISE;][]{wri10}. 
%The infrared data come from the latest survey of the Wide-field Infrared Survey Explorer \citep[WISE;][]{wri10}. The near-infrared data come from new coadds of WISE images and forward model photometry \citep{lan14}.

BASS is a $g-$ and $r$-band imaging survey that uses the wide-field 90Prime camera mounted on the 2.3 m Bok telescope \citep{wil04}. The observation started in 2015.  It mainly covered gray and dark nights from January to July of each year. BASS had its early data and first data releases (DR1) in 2015 December and 2017 January \citep{zou17a}.  DR1 only includes BASS data taken before 2016 July.  By 2017 July,  BASS has completed about 72\% of the whole survey area. The MzLS covers the same area as BASS and have finished more than 76\% of its observations by 2017 July. In this paper, we present the second BASS data release (DR2), which includes both BASS and MzLS data taken as of 2017 July. 

This paper describes the details of BASS DR2. The paper is organized as follows. Section \ref{sec-survey} presents basic information on the BASS and MzLS surveys and their observation status. Section \ref{sec-dr}  updates BASS data reduction. The source detection and photometry are described in Section \ref{sec-phot}. Section \ref{sec-comp} presents the data quality and provides comparisons with other photometric surveys.  Section \ref{sec-data} describes data access and provides guidelines for users. Section \ref{sec-sum} is the summary. 

\section{BASS and MzLS Surveys} \label{sec-survey}
\subsection{Telescope and instruments}
Table \ref{tab-survey} gives a summary of both the BASS and MzLS surveys. BASS uses the Bok telescope to image the northern part of the North Galactic cap with the optical $g$ and $r$ bands. The telescope is located on Kitt Peak near Tucson, Arizona. It is a 2.3 m telescope operated by Steward Observatory of University of Arizona. A wide-field camera, 90Prime, is mounted at the prime focus. It is composed of four 4k $\times$ 4k blue-sensitive CCDs, providing a feild of view (FoV) of 1\arcdeg.08$\times$1\arcdeg.03. The CCD pixel scale is about 0\arcsec.454. There are gaps between the CCDs and the filling factor is about 94\%. 

\begin{table}[!ht]
	\centering
	\tiny
	\caption{A Summary of the BASS and MzLS Surveys}\label{tab-survey}
	  \begin{tabular}{r|cc}
	  \tableline\tableline
	  Telescope and Site & BASS & MzLS  \\
	  \tableline
	  Telescope & 2.3 m Bok & 4 m Mayall \\
	  Site & Kitt Peak & Kitt Peak \\
	  Elevation & 2071 m & 2071 m \\
	  \tableline
	  Camera & BASS & MzLS  \\
	  \tableline
	   Name & 90Prime & Mosaic-3 \\
	   CCD number & 4  & 4\\
	   CCD size & 4096$\times$4032 &4096$\times$4096 \\
	   Gaps & 168{\arcsec} (R.A.), 54{\arcsec} (Decl.) & 52{\arcsec} (R.A.), 62{\arcsec} (Decl.) \\
	   Pixel scale & 0\arcsec.45 & 0\arcsec.26 \\
		FoV & 1\arcdeg.08$\times$1\arcdeg.03 & 36\arcmin$\times$36\arcmin \\
		Gain & 1.5 e/ADU & 1.8 e/ADU\\
		Readout noise & 8.4 e & 9 e\\
		Readout time & 35 s & 30 s\\
		Dark current & 0.8 e/hr & 0.9 e/hr \\ 
	   \tableline
	   Filter & BASS & MzLS  \\
	   \tableline
	   Name & SDSS $g$, DECam $r$ & DECam $z$ \\ 
	   Wavelength & 4789 \AA, 6404 \AA & 9210 \AA \\
	   \tableline
	   Survey parameters & BASS & MzLS  \\
	   \tableline
	   Area & 5400 deg$^2$ ($\delta > 30$\arcdeg) & 5100 deg$^2$ ($\delta > 32$\arcdeg) \\
	   Expected Depth (5$\sigma$) & $g = 24.0$, $r = 23.4$ & $z = 23.0$ \\
	   Observation period & 2015--2018 & 2016--2017  \\
	  \tableline
	  \end{tabular}

\end{table}

MzLS uses the 4 m Mayall Telescope at Kitt Peak to image nearly the same area as BASS with the $z$ band. The telescope is located at the highest peak of the mountain, close to the Bok telescope. Its mirror diameter is about 4 m and its clear aperture diameter is about 3.8 m. The survey uses the Mosaic-3 imager deployed at the prime focus. It is a new wide-field imaging camera installed in mid-2015, replacing the old Mosaic-2 instrument. There are four 4k$\times$4k 500 $\mu$m thick deep-depletion CCDs, which significantly improved the $z$-band observing efficiency. The CCD pixel size is about 0\arcsec.26. The camera FoV is about 36\arcmin$\times$36\arcmin. Each CCD is read out through four amplifiers simultaneously. The readout time in the normal mode is about 30 s. The average gain and readout noise are 1.8 e/ADU and 9 e, respectively. The CCD quantum efficiency is optimized for red wavelengths, which can reach 85\% at 900 nm. The dark current is 0.95 e/hr.  More information about Mosaic-3 can be found in the camera manual\footnote{\url{https://www.noao.edu/kpno/mosaic/manual/}}. 

\begin{figure}[!ht]
\centering
\includegraphics[width=1.0\columnwidth]{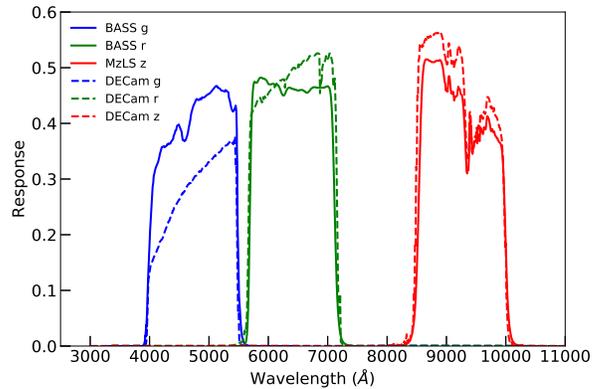}
\caption{Filter response curves for BASS, MzLS, and DECaLS. All curves are the total responses.} \label{fig-filter} 
\end{figure}

In order to achieve homogeneous spectroscopic target selections for DESI, photometric systems of the imaging surveys should be as similar as possible. The DECaLS uses the DECam system, which is also used for the Dark Energy Survey \citep{des05}. The BASS $g$ band is the existing SDSS $g$ filter. It is very close to the DECam $g$ band. The BASS $r$ filter is newly purchased by the Lawrence Berkeley National Laboratory, which is almost the same as the DECam $r$ band. The MzLS uses a newly purchased DECam $z$-band filter. Figure \ref{fig-filter} shows the filter responses of these three filters. The response is the total throughput,  including CCD quantum efficiency, filter transmission, and atmospheric extinction at the zenith, aluminum reflectivity of the prime mirror, and throughput of the optical corrector at the prime focus. The effective wavelengths for the BASS $g$ and $r$ and MzLS $z$ bands are 4789, 6404, and 9210 {\AA}, respectively (see Table \ref{tab-filter} for detailed filter parameters).  For comparison, DECam filter responses are also plotted in Figure \ref{fig-filter} and related parameters are presented in Table \ref{tab-filter}. The Galactic extinction coefficients are $k_g = $3.214, $k_r = $2.165, and $k_z = $1.211, which are adopted from the DECaLS webpage\footnote{\url{http://legacysurvey.org/dr5/description/#galactic-extinction}}.

\begin{table*}[!ht]
\centering
\small
\caption{Parameters for BASS, MzLS and DECaLS filters.} \label{tab-filter} 
 \begin{tabular}{c|c|c|c|c|c|c}
\tableline\tableline
 & BASS $g$ & BASS $r$ & MzLS $z$ & DECam $g$ & DECam $r$ & DECam $z$ \\
 \tableline 
Effective wavelength (\AA) & 4789 & 6404 & 9210 & 4842 & 6439 & 9172 \\
Bandwidth (\AA) & 849 &  837 & 862 & 966 & 895 & 941 \\
FWHM (\AA) & 1435 &  1420 & 1430 & 1290 & 1470 & 1470 \\
 \tableline 
 \end{tabular}
\end{table*}
%$k_\lambda$ &  3.73 &  2.61 & 1.44 & 3.67 &  2.60 & 1.45 \\

\subsection{Survey status}
All DESI imaging surveys have similar observing strategies. Each survey tiles the sky in three passes or exposures. These three passes are dithered to fill CCD gaps. This is also benificial for detecting variable objects and allows zero-point determination with ubercalibration \citep{zou17b}. Pass 1 is observed in both photometric and good seeing conditions. Pass 2 is observed in either photometric or good seeing conditions. Pass 3 is observed in any other usable conditions. The seeing thresholds for BASS and MzLS are 1\arcsec.7 and 1\arcsec.3, respectively. Normally, $g$-band observation occur on dark nights, $r$-band observations occur on dark or gray nights, and $z$ band observations occur on gray and bright nights. The ETC can adjust the exposure time for each pass in real time according to sky brightness, seeing, atmospheric transparency, and Galactic extinction. In this way, both surveys can maintain uniform imaging depths for different passes. 

\begin{figure*}[!ht]
\centering
\includegraphics[width=0.8\textwidth]{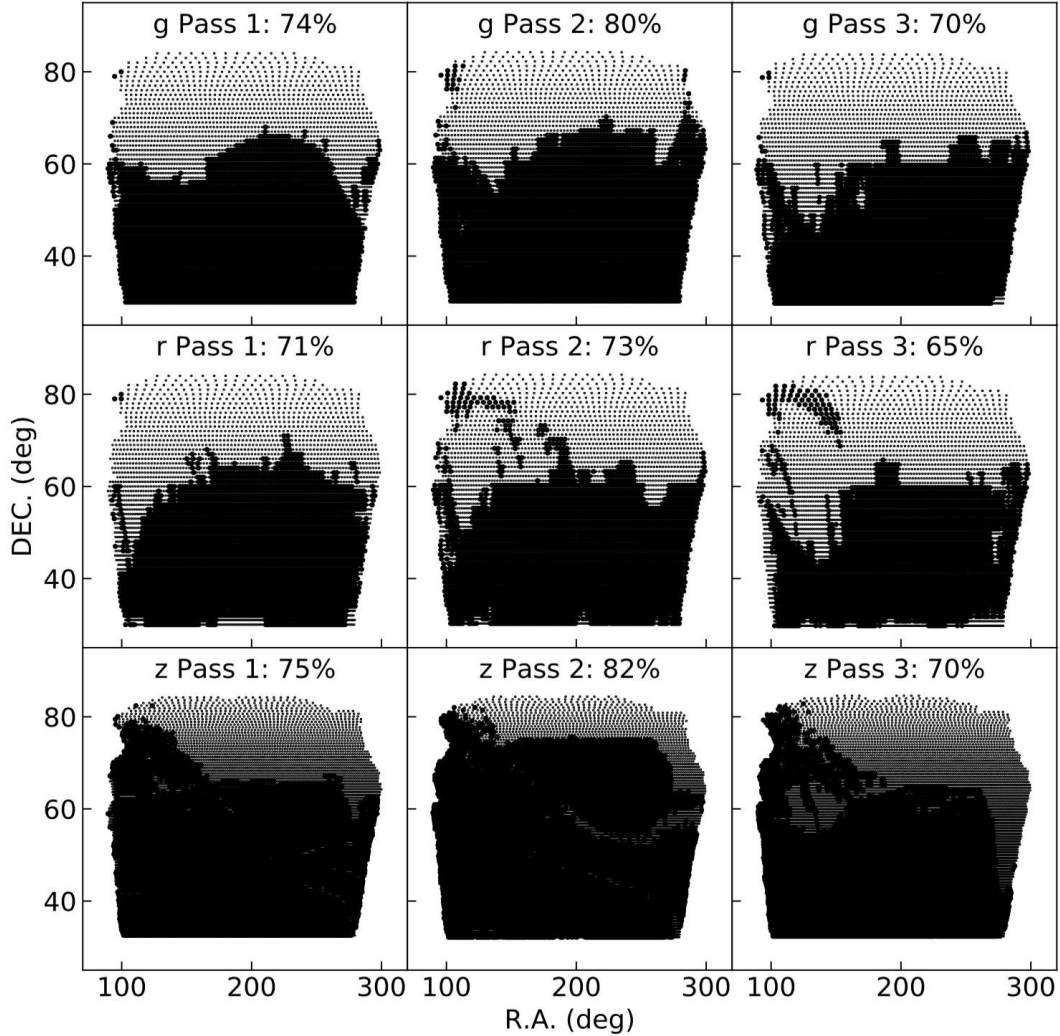}
\caption{Observational progress for BASS and MzLS surveys as of 2017 July for different filters and passes. The percentage of area that has been observed is shown for each filter and pass. }\label{fig-obs}
\end{figure*}

BASS began its observation in 2015 January. The survey was awarded 55, 89, and 92 nights in the spring of 2015, 2016, and 2017, respectively. Since 2015, a number of updates have been implemented, such as instrument control software, telescope flexure maps, and observing tools, which greatly improved the pointing accuracy of the telescope, camera focusing, and observing efficiency. It was discovered that data taken in 2015 suffered from defective electronics in the read-out system, so these data were much noisier. Those electronics were replaced in 2015 September. The MzLS began regular observations in 2016 February. The imaging data suffered from variable crosstalk and pattern noise, which must specially treated. The MzLS has completed its scheduled with about 250 total observations. Figure \ref{fig-obs} presents the observation progress as of 2017 July. Some test regions ($< 100$ deg$^2$) located at Cosmic Evolution Survey (COSMOS) and SDSS Stripe 82 fields are not shown in this figure. The BASS has completed about 72\% of its scheduled observations, while the MzLS has completed about 76\%. With typical observational conditions, BASS is expected to finish all observations in 2018, with an additional 100 nights. The MzLS completed the remaining observations before the Mayall telescope was shut down in 2018 February. 

\section{Updates on Data Reduction} \label{sec-dr}
\subsection{Image processing}
The BASS and MzLS raw data are processed using the same pipeline. The basic corrections of overscan, bias, flat, and crosstalk effect are almost the same as those in BASS DR1. We make minor modifications when dealing with BASS data taken in 2015. The read-out mode was different in 2015. Bias frames were taken sequentially without flushing. Consequently, only the first bias frame can be used for bias subtraction. In addition, gain, readout noise, and crosstalk coefficients for each amplifier are calculated after the camera was remounted every summer. The identification of cosmic rays is also different in DR2. A Laplacian algorithm\footnote{\url{https://pypi.python.org/pypi/astroscrappy/1.0.3}} originating from \citet{van01} was used to detect cosmic rays.

\begin{figure*}[!ht]
\centering
\includegraphics[width=0.8\textwidth]{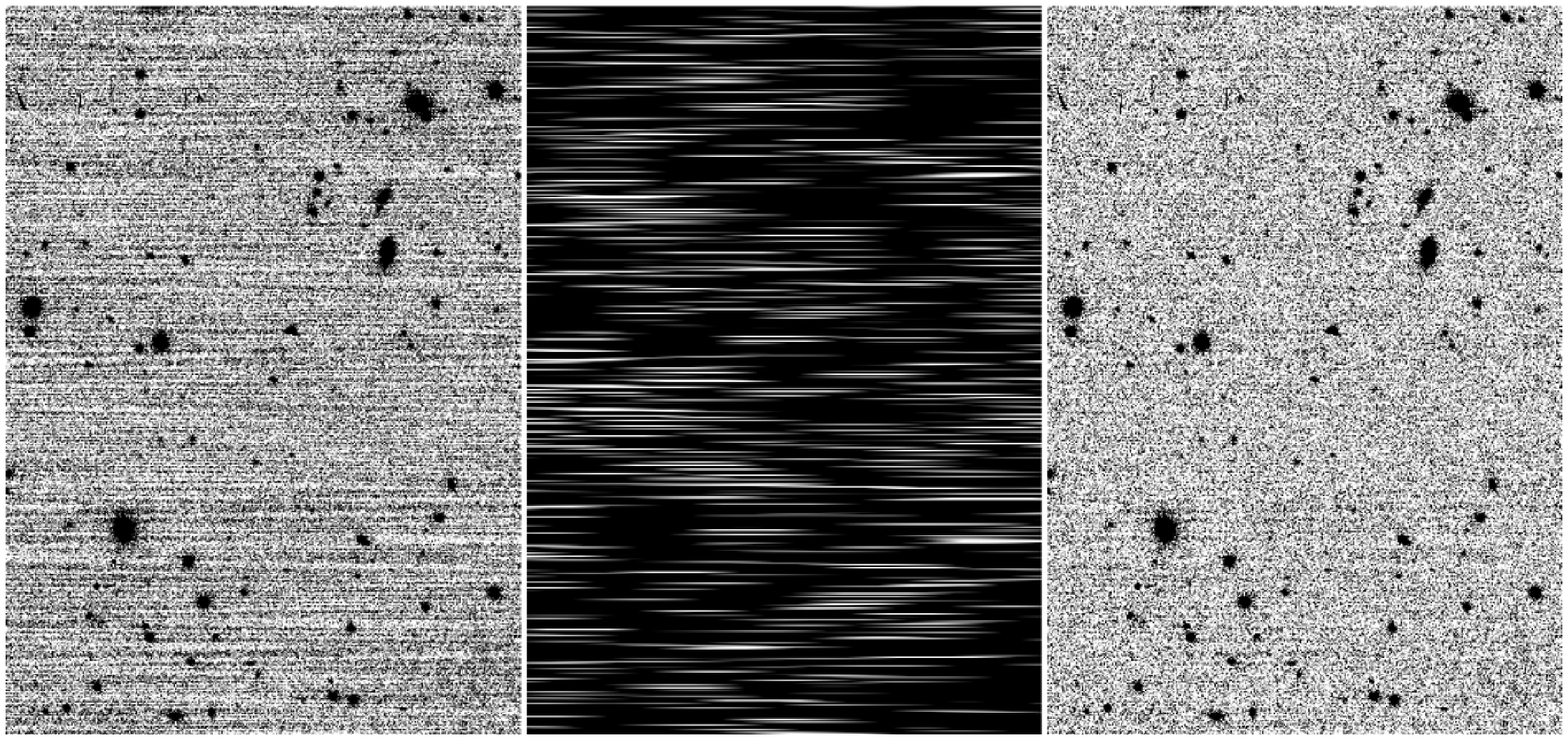}
\caption{Removal of pattern noise in MzLS data. Left: raw image; middle: noise pattern constructed by a low-pass filter; right: image after pattern removal.}\label{fig-noise}
\end{figure*}

The crosstalk effect in MzLS data is much more serious than that in BASS data. The maximum intra-CCD crosstalk coefficient in MzLS reaches up to 5.2\%, while for BASS it is only about 0.3\%. There is also serious pattern noise in MzLS caused by the electromagnetic interference during CCD reading out, which is demonstrated in the left panel of Figure \ref{fig-noise}. We apply a low-pass filter to reconstruct the noise pattern. First, a 5-order low-pass Butterworth digital filter is designed by using the ``butter" program in a Python package  of SciPy \footnote{https://docs.scipy.org/doc/scipy/reference/generated/scipy.signal.butter.html}. Then, the low-pass filter is applied to the source-removed $z$-band image and a smoothed pattern is constructed as shown in the middle panel of Figure \ref{fig-noise}. Finally, the pattern is subtracted from the original image as shown in the right panel of this figure. 

\subsection{Astrometry}
The Software for Calibrating AstroMetry and Photometry \citep[SCAMP;][]{ber06} is used for calculating astrometric solutions. In DR1, the SDSS and Two Micron All Sky Survey \citep[2MASS;][]{skr06} point-source catalogs were used as the reference catalogs. However, the Gaia \citep{gai16b} catalog is adopted in DR2. The first data release of Gaia \citep[hereafter Gaia DR1; ][]{gai16a} was published in 2016 September. It provides accurate astrometric measurements for objects in the $G$ band down to 20.5 mag over the entire sky. There are some images (about 1.5\%) with faulty astrometry due to sparse sources in $Gaia$ catalogs, the 2MASS catalog is used to derive astrometric solutions for these. 

\begin{figure*}[!ht]
\centering
\includegraphics[width=\textwidth]{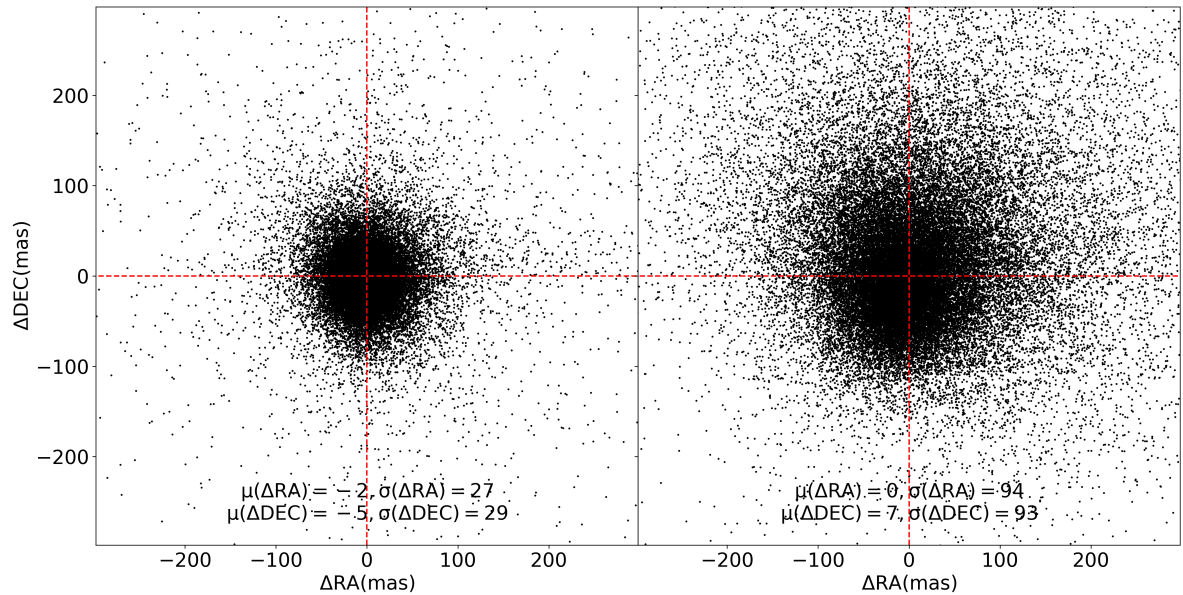}
\caption{Astrometric accuracy using Gaia DR1 (left) and SDSS DR9 (right) as the reference catalogs. The data points are the same bright objects. The median offsets and RMS errors in R. A. and decl. are also shown in each panel. The horizontal and vertical dashed lines show zero offsets.}  \label{fig-astrom}
\end{figure*}

Figure \ref{fig-astrom} gives a comparison of the astrometric accuracy when using different reference catalogs. The astrometric accuracy is much improved when using the $Gaia$ DR1 catalog as the reference. The median errors in R. A. and decl. are both about 28 mas. By contrast,  if the SDSS catalog is used, the median errors in R. A. and decl. are about 93 mas.

\subsection{Zero-point calculation}
As in DR1, the DR2 still uses Pan-STARRS1 data \citep[hereafter PS1;][]{cha16} to derive photometric solutions.  A large aperture diameter of 26 pixels is used for counting source fluxes. The aperture is equivalent to about 11\arcsec.9 for BASS and 6\arcsec.8 for MzLS (i.e. about 7 times the seeing FWHM). Point sources with S/N larger than 10 are selected and then matched to the PS1 catalogs with a cross-matching radius of 2\arcsec. The PS1 magnitudes are transformed to the BASS and MzLS photometric systems using the following equations:
\begin{figure*}
\begin{align} 
    (g-i) &= g_{\rm PS1}-i_{\rm PS1},\\
    g_{\rm BASS} &= g_{\rm PS1} - 0.08826 + 0.10575(g-i)  - 0.02543(g-i)^2 + 0.00226(g-i)^3, \\ 
    r_{\rm BASS} &= r_{\rm PS1} + 0.07371 - 0.07650(g-i) + 0.02809(g-i)^2 - 0.00967(g-i)^3, \\
    z_{\rm MzLS} &= z_{\rm PS1} + 0.10164 - 0.08774(g-i) + 0.03041(g-i)^2 - 0.00947(g-i)^3.
    \label{equ-colorterm}
\end{align}
\end{figure*}
These color terms are valid for stars with $0.4 < g-i < 2.7$. The constant terms are kept to ensure no systematic offsets between BASS/MzLS and PS1. The zero-point of a CCD image is calculated as the error-weighted difference between instrumental aperture magnitudes and transformed PS1 PSF magnitudes with outliers removed.

\begin{figure}[!ht]
	\centering
	\includegraphics[width=\hsize]{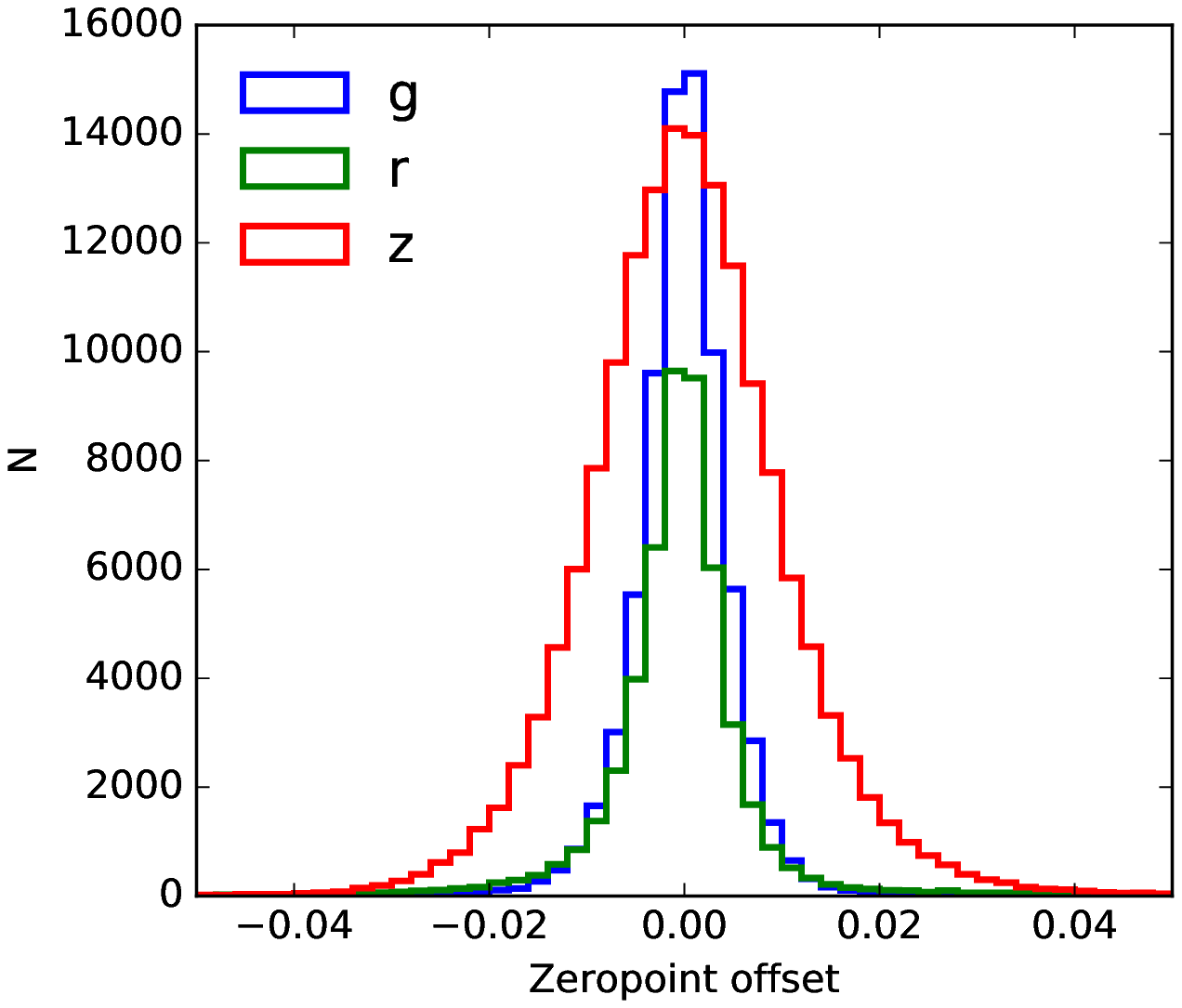}
	\caption{Distributions of internal zero point corrections for $g$ (blue), $r$ (green), and $z$ (red) bands. }
	\label{fig-zpoff}
\end{figure}

\begin{figure}[!ht]
	\centering
	\includegraphics[width=\hsize]{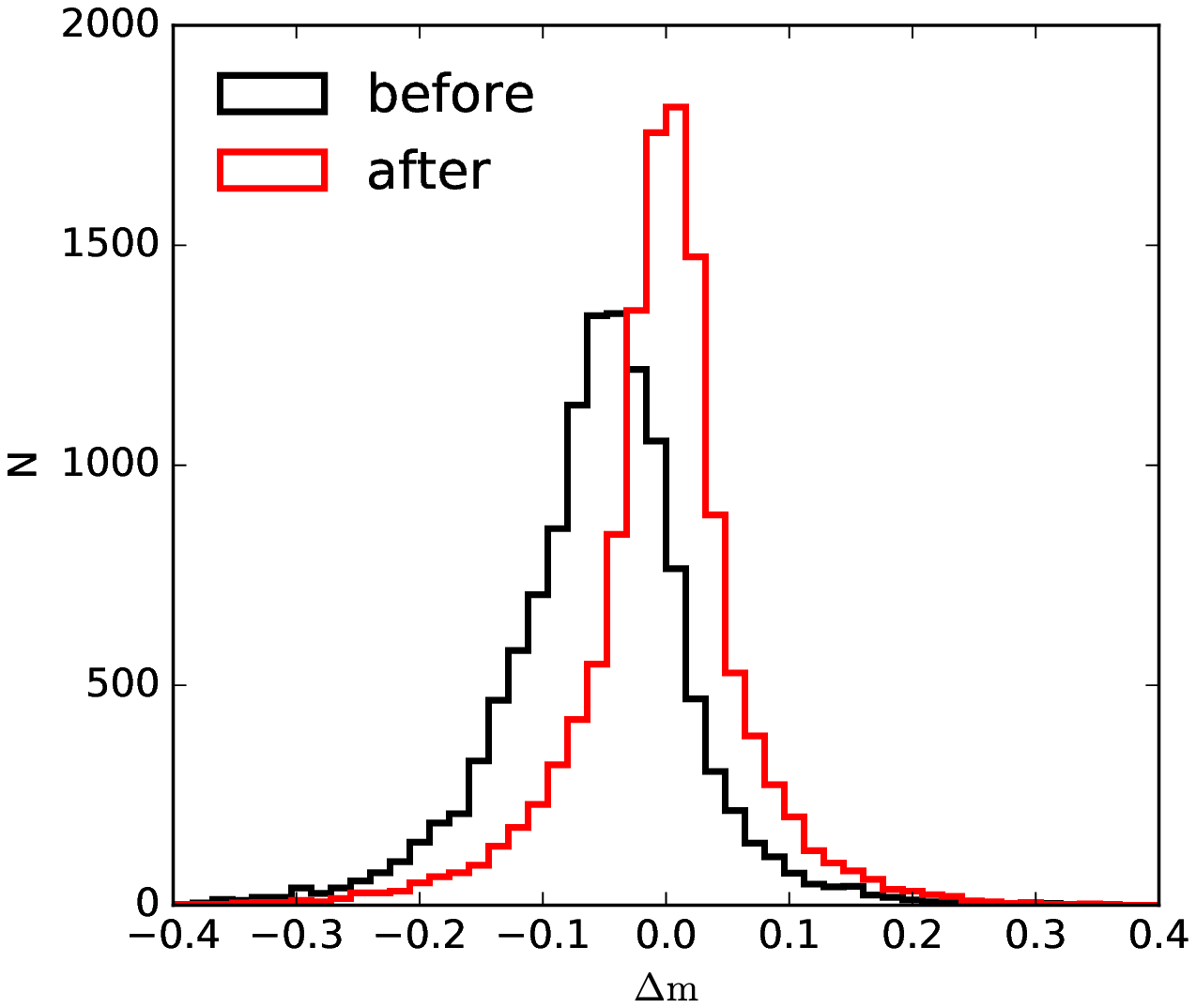}
	\caption{Distributions of magnitude differences of common objects in a $g$-band image and its adjacent images before (black) and after (red) internal zero point corrections are applied. }
	\label{fig-zpmagdiff}
\end{figure}

In addition to the zero-point obtained with the external catalogs discussed above, we can also calculate an internal zero-point by using objects located in overlaps of different exposures.  We have three passes for each field.  Each pass has sufficiently large position offsets.  For a specified image, an internal zero-point offset is calculated by comparing magnitudes of common objects in adjacent images. This offset is derived iteratively until its change is less than 0.001. We require that there are at least 50 stars to calculate the offset and the maximum number of iterations is 20. Normally, the computations for most images are converged after 10 steps.  Figure \ref{fig-zpoff} shows the distributions of the zero-point offsets for different filters. The dispersions are 0.005, 0.008 and 0.01 for $g$, $r$, and $z$ bands, respectively. Figure \ref{fig-zpmagdiff} gives an example of the performance of zero-point correction in one specially selected $g$-band image. This image is out of the PS1 coverage, so we use the Fourth US Naval Observatory CCD Astrograph Catalog to calculate a crude external zero-point.  The histograms in this figure show the distributions of magnitude differences of common objects in multiple exposures before and after zero-point corrections are applied.  We can see that after applying the corrections, both offset and scatter decrease. In practice, the internal zero point is used in DR2 only if the number of common stars are larger than 100 and its RMS error is less than 0.1. Otherwise the external zero-point is used.  

\section{Photometry} \label{sec-phot}
There are significant changes in source detection and photometry between DR1 and DR2. In DR1,  the source detection and corresponding photometry are performed on single-epoch images. However, we detect sources in stacked images in order to improve the detection efficiency in DR2. The DR2 photometry occurs in single-epoch images based on the positions of those sources.

\subsection{Image stacking and source detection} \label{sec-detection}
As described in \citet{zou17a}, the full sky is equally divided into 96,775 blocks. These blocks are evenly spaced in decl.. Each block has an area of 0\arcdeg.681$\times$0.681 deg$^2$, equal to an image size of 5400 $\times$ 5400 with a pixel scale of 0\arcsec.454. There are overlaps of 0\arcdeg.02 in both R.A. and decl. between adjacent blocks. The stacked image is generated with a simple tangent-plane projection around the block center.

\begin{figure*}[!ht]
\centering
\includegraphics[width=0.8\textwidth]{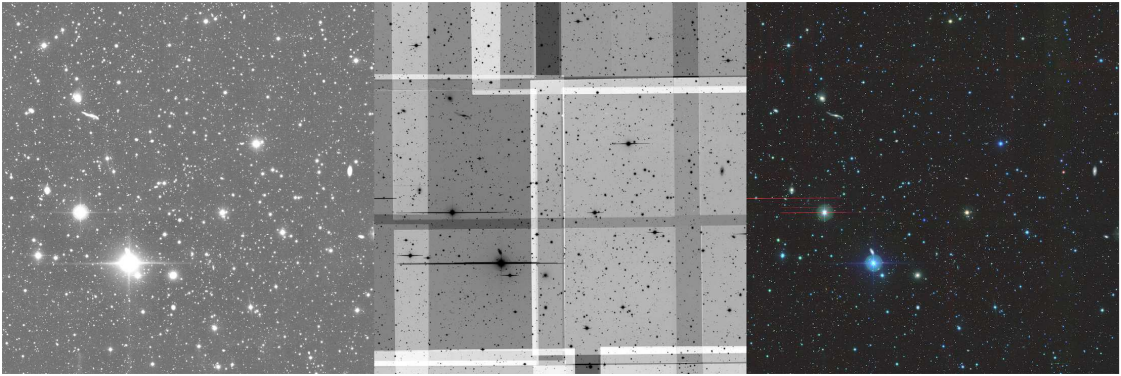}
\caption{Left: a $g$-band stacked image of an arbitrary block. Middle: corresponding weight image. Right: color image combined from $g$, $r$, and $z$-band stacked images.}\label{fig-stack}
\end{figure*}

We create four stacked images: three images for the $g$, $r$, and $z$ bands and one composite from these three stacks. The composite image is combined from $g$, $r$, and $z$-band stacked images with flux scales of 0.65, 1.0, and 1.5, respectively, which makes the $g - r$ and $r - z$ colors of F/G type stars close to zero. These flux scales are also used for generating color pictures. For a specified band and block, we collect single-epoch images that are connected to the block. These images are resampled and reprojected by Swarp \footnote{\url{http://www.astromatic.net/software/swarp}} and then combined by median to form a stacked image and corresponding weight image \citep{ber02}. The sky background map for each single-exposure image is estimated by using a mesh with a grid size of 410 pixels and masking large objects from the Third Reference Catalog of bright galaxies (RC3) and New General Catalog (NGC). The flux of each single-epoch image is scaled to make the stack image have a photometric zero-point of 30 (i.e. magnitude is calculated as $m = -2.5\mathrm{log}_{10} F + 30$). The single-epoch images used for stacking should satisfy the following conditions: (a) exposure time $>$ 30 s; (b) seeing $<$ 3\arcsec.5 for BASS and $<$ 2\arcsec.5 for MzLS; (c) zero-point RMS error $<$ 0.2; (d) number of stars used for calculating the zero-point $>$ 50; (e) astrometric RMS error in both R.A. and decl. $<$ 0\arcsec.5; (f) 5$\sigma$ depths are at most 1.5 mag shallower than the required one; (g) sky ADU $<$ 15,000 for BASS and $<$ 25,000 for MzLS; (h) transparency $>$ 0.5. Here the transparency is related to the level of atmospheric extinction. It is defined as $10^{0.4(\mathrm{zp_0}-\mathrm{zp})}$, where zp is the zero-point of an image corrected with airmass and $\mathrm{{zp}_0}$ is the typical zenith zero point on clear nights. Figure \ref{fig-stack} gives examples of a $g$-band stacked image, its weight map, and corresponding color image composed of three-band stacked images. 

Source detections are performed first on the composite image and then separately on $g$-, $r$-, and $z$-band stacked images by SExtractor \citep{ber96}. Objects are kept only if they are detected at least twice in those four images. In this way, a majority of fake sources with brightnesses close to the detecting threshold in each stacked image can be filtered out. Table \ref{tab-det} lists some key configuration parameters that significantly impact our detection and deblending. We chose a relative small threshold, minimum number of connected pixels, and convolution kernel so as to detect as many objects as possible for a single band. In addition, the minimum contrast parameter of ``DEBLEND\_MINCONST" is set to 0.001, which is determined as a compromise. A higher value will lead to more fragments of a large object and more fake sources within this object, while a lower value will give fewer fragments.

\begin{figure*}[!ht]
\centering
\includegraphics[width=1.0\textwidth]{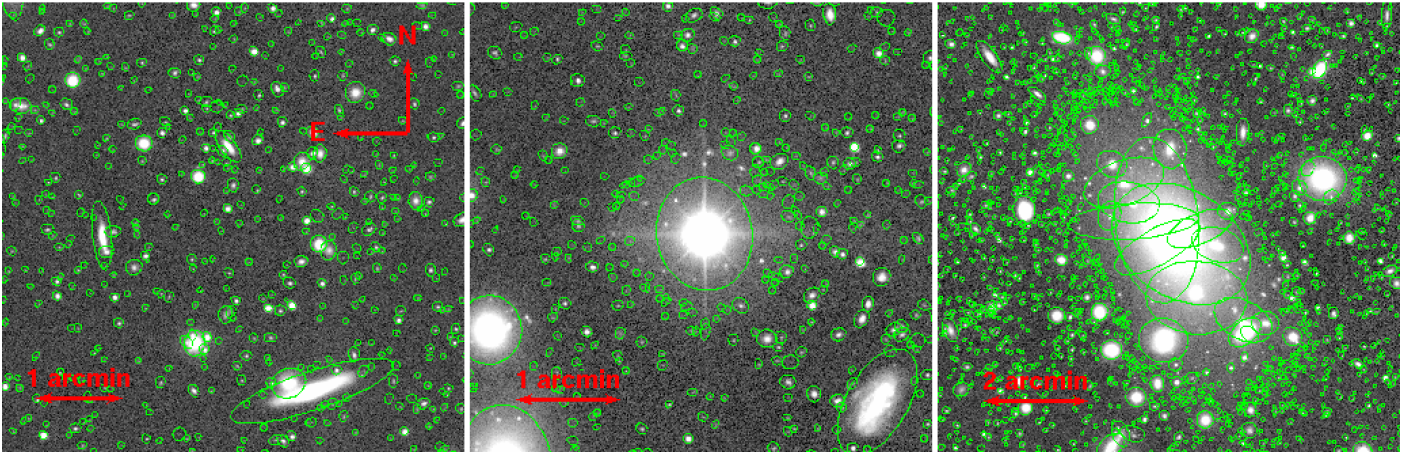}
\caption{Detected sources overlain with Kron elliptical apertures . Left: normal galaxies and stars. Middle: detections around a bright star. Right: detections around a large galaxy. These images are displayed in a asinh scale to highlight the outer parts of sources more clearly.  }\label{fig-detection}
\end{figure*}

Figure \ref{fig-detection} shows Kron elliptical apertures (see Section \ref{sec-kron}) of final detected sources on composite images. From the left and middle panels of this figure, we can see that the deblending algorithm is suitable for small and medium-sized galaxies (typical size of about 1 arcmin). For larger extended galaxies, the deblending method may not be appropriate due to bright foreground stars and substructures hesitated in these galaxies. We also notice that there are still some fake sources surrounding large bright stars or galaxies that are detected due to large noise fluctuation. As checked with external deep catalogs from COSMOS and DEEP2\footnote{\url{http://deep.ps.uci.edu/DR4/photoprimer.html}}, about 8\% of objects might be spurious. Most of these objects are located around bright stars and large galaxies. A few of them are probably CCD artifacts such as cosmic rays (especially in the $z$ band), which are not well identified by the imaging pipeline or not effectively removed due to a lack of enough exposure numbers.

\begin{table*}[!ht]
\centering
\footnotesize
\caption{Key configuration parameters for source detection in SExtractor} \label{tab-det} 
 \begin{tabular}{r|c|l}
\tableline\tableline
Parameter & Value & Description  \\
 \tableline 
DETECT\_MINAREA  &  3                &      Minimum number of pixels above threshold \\
THRESH\_TYPE        &  RELATIVE   &     Threshold type \\
DETECT\_THRESH    &  1.0     &  Detection threshold above the local background \\
FILTER\_NAME          & gauss\_1.5\_3x3.conv   & Name of the file containing the filter for detection \\
CLEAN                     &   N                                & Whether to clean spurious detections \\
DEBLEND\_NTHRESH  & 64                             & Number of deblending sub-thresholds \\
DEBLEND\_MINCONT  & 0.001                        & Minimum contrast parameter for deblending \\
\tableline 
\end{tabular}
\end{table*}

\subsection{Methods of photometry} 
We have developed a new Python-based package for photometry (H. Zou et al. in preparation). It can be used for more general purposes, but at present it is just applied to BASS DR2. The current code can make accurate measurements of circular aperture, elliptical aperture, and PSF magnitudes at specific positions. However, model measurements with galaxy profiles, such as deVauculeurs, exponential profiles, and composite profiles, are still in development.  We describe a few key features below. 

\subsubsection{Segmentation and shape measurements}
Different kinds of magnitude measurements are performed on a single-epoch image for objects detected on stacked images. These objects are first projected onto the image. A watershed segmentation algorithm \footnote{\url{http://scikit-image.org/docs/dev/auto_examples/segmentation/plot_watershed.html}} is then adopted to separate signal pixels belonging to each object. The signal pixels are identified with a detecting threshold at 1$\sigma$ above the sky background after the image is smoothed with a Gaussian kernel of $\sigma$ = 1.5 pixels. The global sky background and its RMS map are calculated in mesh grids. Based on the segmentation of each object, we calculate the centroid, shape parameters, and refined center using a Gaussian-kernel window. The shape parameters describe the shape of an object as an ellipse, including semi-major ($A$) and semi-minor ($B$) axis lengths and position angle (PA). They are also called as $1\sigma$ elliptical parameters. 

\subsubsection{Circular aperture photometry}
We adopt 12 apertures for circular aperture photometry with radii ranging from 3 to 40 pixels. These aperture sizes are the same as those used in the South Galactic Cap $u$-band Sky Survey, which utilized the Bok telescope and the 90Prime camera to perform a $u$-band imaging survey \citep{zou15,zou16, zho16}. Table \ref{tab-aperture} shows the aperture radii in both pixels and arcseconds.
\begin{table*}[!ht]
\centering
\scriptsize
\caption{Aperture Radii Used for Circular Photometry.} \label{tab-aperture} 
 \begin{tabular}{r|c|c|c|c|c|c|c|c|c|c|c|c}
\tableline\tableline
Aperture No. & 1 & 2 & 3 & 4 & 5 & 6 & 7 & 8 & 9 & 10 & 11 & 12 \\
 \tableline 
Radius in pixel & 3 & 4 & 5 & 6 & 8 & 10 & 13 & 16 & 20 & 25 & 30 & 40 \\
Radius in arcsec (BASS) & 1.36 & 1.82 & 2.27 & 2.72 & 3.63 & 4.54 & 5.90 & 7.26 & 9.08 & 11.35 & 13.62 & 18.16 \\
Radius in arcsec (MzLS) & 0.78 & 1.04 & 1.31 & 1.57 & 2.09 & 2.61 & 3.39 & 4.18 & 5.22 & 6.53 & 7.83 & 10.44  \\
 \tableline 
 \end{tabular}
\end{table*}

\subsubsection{Isophotal and Kron elliptical aperture photometry} \label{sec-kron}
Isophotal magnitudes are derived by simply integrating pixel fluxes within segments. For better magnitude measurements of galaxies, we estimate an appropriate elliptical aperture for each object. The elliptical aperture should enclose most flux. We use the ``Kron aperture", which was introduced in \citet{kro80}. The Kron aperture size is described by the Kron radius. This radius is determined in a similar way as in SExtractor.  First, a characteristic radius $r_1$ is calculated as the first-order moment within an large ellipse, whose size is 6 times the $1\sigma$ ellipse: $r_1 = \frac{\sum{rF(r)}}{\sum{F(r)}}$, where $r$ is the elliptical distance of a pixel to the center and $F(r)$ is the corresponding flux in this pixel. We set the Kron radius $r_k$ to be 2.5$r_1$. The Kron aperture magnitudes are measured by integrating the fluxes of pixels within the Kron ellipse. The major and minor axis lengths of the Kron ellipse are computed as $r_k\sqrt{e}$ and $\frac{r_k}{\sqrt{e}}$, respectively, where $e$ is the elongation. To avoid extremely small and large apertures, we set lower and upper limits of the Kron radius to be 3$r_0$ and 10$r_0$, where $r_0 = \sqrt{AB}$. It is shown that more than 94\% of the light for galaxies is located in the Kron ellipse, almost independent of their magnitudes. For both circular and Kron aperture photometry, special handling is done for objects contaminated by nearby sources in order to improve the photometric accuracy. In the aperture, the pixels occupied by nearby objects are filled by mirroring the opposite pixels relative to the object center. In addition, each pixel is divided into 5$\times$5 sub-pixels to more accurately count the fluxes of pixels at the boundary of the aperture.

\subsubsection{PSF photometry}
PSF magnitudes are measured with the PSF model derived by PSFEx \citep{ber11}. PSFEx models the PSF as a linear combination of basis vectors. The pixel basis and automatic sampling step are selected in this software. We use a third-degree polynomial to model the position-dependent variation of the PSF. The PSF size is 45$\times$45 for both BASS and MzLS, which is about 12 times the seeing FWHM. Table \ref{tab-psfex} lists some key input parameters in PSFEx. When fitting with the PSF model using our code, the pixels belonging to the segment of each object are used. If objects are not isolated, i.e. their segments are connected to each other, they are fitted simultaneously. We use k-means clustering to iteratively group objects. First, two groups are generated by the k-means clustering algorithm. Then the members in each group are divided into two subgroups using the same algorithm. All groups are divided in this way until the number of members in each group is not larger than three. Finally, all groups are ranked by their brightest members, and the PSF magnitudes in each group are measured simultaneously.

\begin{table*}[!ht]
\centering
\footnotesize
\caption{Key input parameters for PSFEx} \label{tab-psfex} 
 \begin{tabular}{r|c|l}
\tableline\tableline
Parameter & Value & Description  \\
 \tableline 
BASIS\_TYPE         &    PIXEL\_AUTO       & Type of basis vector \\
PSF\_SAMPLING       &    0.0               & Sampling step in pixel units \\
PSF\_ACCURACY       &    0.01              & Accuracy to expect from PSF "pixel" values \\
PSF\_SIZE           &    45, 45             & Image size of the PSF model \\
PSF\_RECENTER       &      Y               & Whether to allow recentering of PSF-candidates \\
CENTER\_KEYS        &  X\_IMAGE, Y\_IMAGE   & Catalog parameters for source pre-centering \\
PSFVAR\_DEGREES	    &     3                & Polynomial degree for position-dependent variation \\
SAMPLE\_FWHMRANGE   &   2.0, 20.0           & Allowed FWHM range \\
SAMPLE\_VARIABILITY &   0.5                & Allowed FWHM variability \\
SAMPLE\_MINSN       &    20                & Minimum S/N  \\
SAMPLE\_MAXELLIP    &    0.3               & Maximum ellipticity  \\
 \tableline 
 \end{tabular}
\end{table*}

\subsection{Co-adding measurements}
Based on the detections on stacked images, we perform circular aperture, isophotal, Kron elliptical aperture, and PSF flux measurements on single-epoch images. As described in our DR1 paper \citep{zou17a}, the coordinates of objects are corrected with astrometric residuals, which mainly origin from poor charge transfer efficiencies of detectors. The amount of corrections is about a few percent of an arcsecond. All magnitudes are aperture-corrected with growth curves produced by circular aperture photometry. In addition, magnitudes are corrected with photometric residuals, which are mainly caused by the focal distortion, improper flat-fielding, and scattered light. These astrometric and photometric residuals are obtained by comparing the coordinates and magnitudes with those in the Gaia DR1 and PS1 catalogs, respectively. 

The parameters measured in single-epoch images are merged to generate co-added catalogs. For each object, we take weighted averages of the refined centers, shape parameters, and fluxes (equivalently magnitudes). Here, the weight is the inverse variance ($w=\frac{1}{\sigma^2}$). Thus, the errors of these averages come from errors of single-epoch measurements, which are computed as $\frac{1}{\sqrt{\Sigma w_i}}$, where $w_i$ is the weight. We also calculate the parameter standard error, which is rms of multiple parameter measurements divided by the square root of the number of measurements. The standard error including all possible error sources is a more realistic error estimation. Particularly, a large magnitude standard error for a variable star indicates light variation. The effective seeing and sky background, number of exposures, and average, minimum, and maximum of the Julian day when observations were taken, are recorded in the catalogs.

\section{Data Analyses and Comparisons} \label{sec-comp}
\subsection{Data quality of single-epoch images}
The ETC ensures our imaging depths to meet the requirements, although the observing strategy entails different passes being observed under different weather conditions. Table \ref{tab-quality} presents median observational and imaging parameters for different filters. The $z$-band seeing is the best. The depth is estimated as the median PSF magnitude at the error of about 0.21 mag (5$\sigma$). The single-epoch depths are 23.3, 23.1, and 22.6 mag for the $g$, $r$, and $z$ bands, respectively. We also have checked that the depths for different passes of the same filter are quite similar owing to the ETC. 
\begin{table*}[!ht]
\centering
\caption{Median observational and imaging parameters for single-epoch images} \label{tab-quality}
\begin{tabular}{c|cccccc}
\hline
\hline
\multirow{2}{*}{\backslashbox{Filter}{Parameter}} & Area  & Seeing & Sky & Airmass & Zero-point & Depth \\
                                                                           &  (a) & (b) & (c)  & (d) & (e) & (f) \\
\hline  
 $g$ & 75\% & 1.60  & 22.09  & 1.05 & 25.92 & 23.33 \\
 $r$ &  70\% & 1.42  & 21.27  & 1.05 & 25.81 & 23.06 \\
 $z$ &   76\%  & 1.04 & 18.96   & 1.05  &  26.46  & 22.59 \\                                                          
\hline
\end{tabular}
\tablenotetext{}{Notes: (a) percentage of tiles that have been observed; (b) seeing in arcseccons calculated from the FHWMs of stars; (c) sky brightness in mag arcsec$^{-2}$; (d) airmass of the telescope pointing; (e) zero-point in AB mag for 1 e/s; (f) 5$\sigma$ depth in mag estimated by the PSF magnitude.}
\end{table*}     

\subsection{Co-added Depths}
\begin{figure}[!ht]
\centering
\includegraphics[width=\columnwidth]{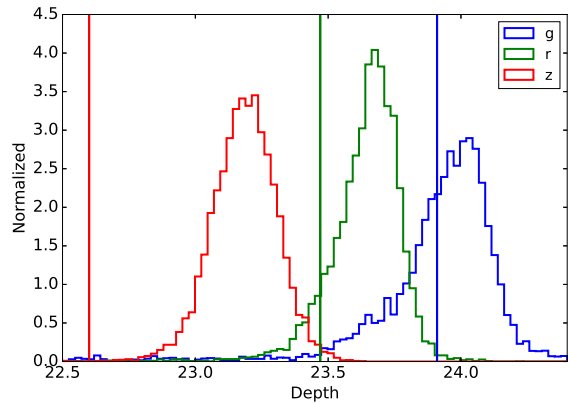}
\caption{Normalized distributions of full depths estimated by $5\sigma$ PSF magnitude errors for the $g$, $r$, and $z$ bands. The vertical lines indicate the DESI required depths.}\label{fig-depth}
%The solid histograms represent for the depths estimated by the measuring error and the dashed ones represent for the depths estimates by the standard error. 
\end{figure}

The depth requirements for DESI are 23.91, 23.47, and 22.60 mag for the $g$, $r$, and $z$ bands, respectively, which are defined by 5$\sigma$ extended sources with a correction of Galactic extinction and converted to our photometric systems with color terms as shown in Equation (\ref{equ-colorterm}). The Galactic reddening map comes from \citet{sch98}. We estimate the $5\sigma$ depths by using co-added PSF magnitude errors. Figure \ref{fig-depth} shows the depth distributions of all blocks with exposure numbers equal to 3. They are so-called ``full depth," with three complete passes. The $g$-band distribution has a fatter tail at the bright end. Because of this we include the data taken in 2015. These data were shallower due to the imperfect ETC and problematic due to bad A/D converters. The low bits of ADU integers were lost. This seriously affects the photometry of faint sources. This issue mostly affects the $g$-band images. Some of the affected areas have been re-observed in recent years and will be totally covered in future observations. Table \ref{tab-depth} lists the median depths and fractions of objects for different exposure numbers. The median 5$\sigma$ depths for the $g$, $r$, and $z$ bands are 24.05, 23.61, and 23.10 mag, respectively. Most of the area covered by the $g$ band has more than 3 exposures due to duplicate observations from 2015. 

\begin{figure}[!ht]
\centering
\includegraphics[width=\hsize]{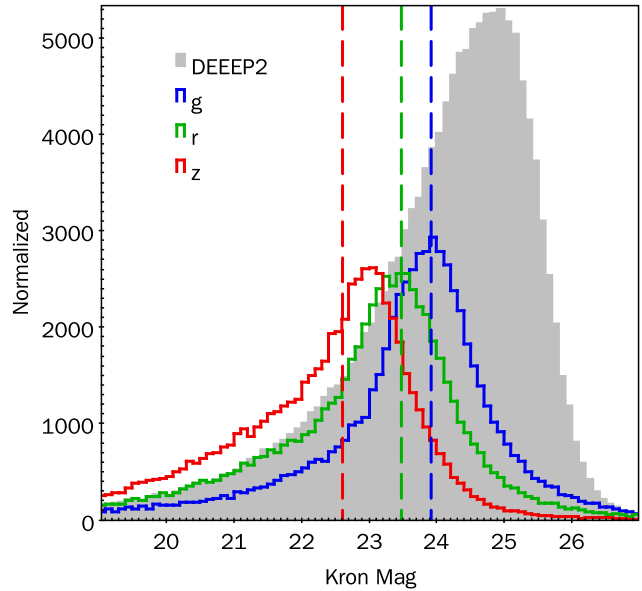}
\caption{Kron magnitude distributions of sources located in DEEP2 fields. The nominal DESI depths are plotted as dashed vertical lines. The filled gray histogram shows the DEEP2 $R$-band magnitude distribution after it is converted to the BASS $r$ band using a system transformation.}\label{fig-nc}
\end{figure}

\begin{figure*}[!ht]
\centering
\includegraphics[width=0.8\textwidth]{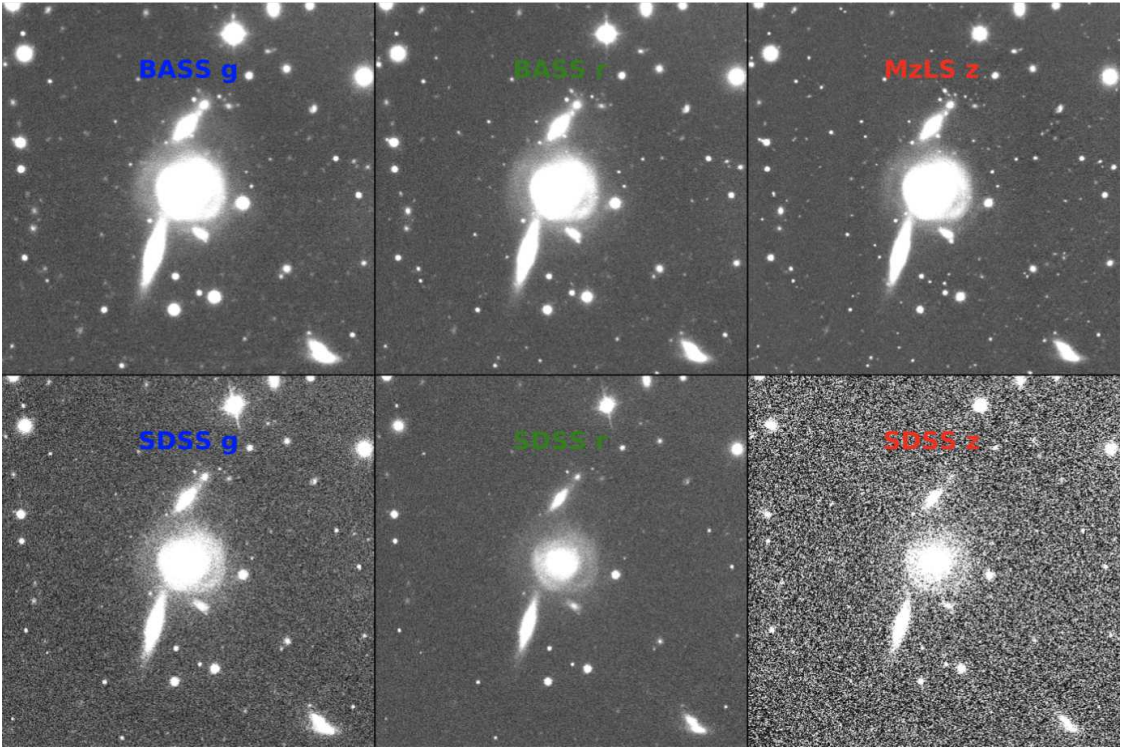}
\caption{Imaging comparison between BASS/MzLS and SDSS. The scale for each band is the same. }\label{fig-imagecomp}
%The solid histograms represent for the depths estimated by the measuring error and the dashed ones represent for the depths estimates by the standard error. 

\end{figure*}

Figure \ref{fig-nc} shows the number counts of objects in the $g$, $r$ and $z$ bands in the regions of DEEP2 fields. The Kron magnitude is adopted here, which is a good brightness measurement for both stars and galaxies. The DEEP2 photometric data were taken with the CFH12K camera on the 3.6 m Canada-France-Hawaii Telescope and cover about 5 deg$^2$ over the sky, with completeness limits of $B = 25.25$, $R = 24.75$, and $I = 24.25$ mag \citep{coi04}. We overplot the $R$-band number count in Figure \ref{fig-nc} after the $R$-band magnitude is converted to BASS $r$ using a color term based on $B-R$. In general, the BASS and MzLS completeness limits reach or exceed the nominal DESI depths and the $r$-band magnitude distribution matches well with the DEEP2 $R$-band distribution. About 2\% of objects have BASS $r$-band exposure numbers less than 3, so corresponding magnitude limits are about 0.3--0.6 mag shallower. This causes a slightly lower number count at $r \sim 22$. In addition, different photometric systems and photometric methods between BASS and DEEP2 might also affect the discrepancy of the magnitude distribution. The co-added depths are 1.5--2.5 mag deeper than the SDSS imaging. For a visual comparison, we demonstrate the imaging data around a group of galaxies from both BASS/MzLS and SDSS in Figure \ref{fig-imagecomp}. There are many more fainter objects that can be seen in BASS/MzLS images. The SDSS $z$-band is much shallower, leading to a much noisier background. 

\begin{table*}[!ht]
\centering
	\caption{Imaging depths and fractions of objects with different exposure numbers}\label{tab-depth}
	  \begin{tabular}{c|ccc|ccc}
	  \tableline\tableline
	               & \multicolumn{3}{c|}{Median Depth}  &  \multicolumn{3}{c}{Fraction}  \\
	  \tableline
	  \backslashbox{Exposure}{Filter} & $g$ & $r$ & $z$ & $g$ & $r$ & $z$ \\
          \tableline
	  1  & 23.31 & 22.99 & 22.63 &  5.7\% & 9.6\% & 13.7\%\\
	  2  & 23.74 & 23.44 & 22.98 &  11.9\% & 18.6\% & 28.5\% \\
	  3  & 23.91 & 23.64 & 23.18 &  18.1\% & 43.2\% & 38.6\%\\
	  $\geq 4$& 24.12 & 23.80 & 23.38 & 64.3\% & 28.6\% & 19.2\%\\
	  median & 24.05 & 23.61 & 23.10 & \nodata & \nodata & \nodata \\
	\tableline
	  \end{tabular}
\end{table*}

\subsection{Photometric comparisons}
\begin{figure*}[!ht]
\centering
\includegraphics[width=0.8\textwidth]{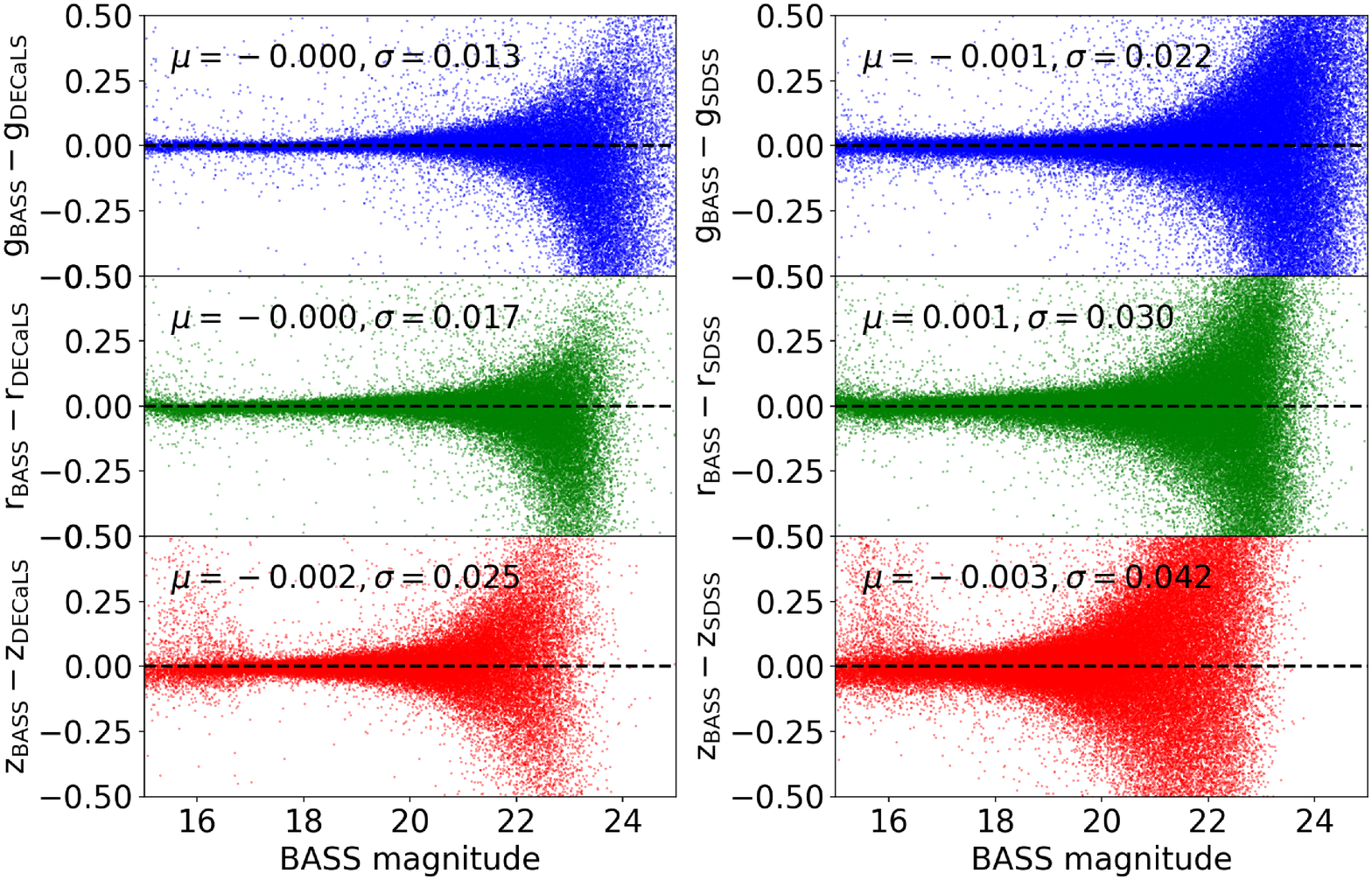}
\caption{Magnitude difference as a function of BASS magnitude between BASS and DECaLS (left panels) and between BASS and SDSS (right panels). The $g$, $r$, and $z$ bands are plotted in blue, green, and red in the top, middle and bottom panels, respectively. The magnitude average offset and scatter of stars brighter than 20 mag are shown in each panel. The horizontal dashed lines show zero offsets.}\label{fig-magcomp}
\end{figure*}

\begin{figure*}[!ht]
\centering
\includegraphics[width=0.8\textwidth]{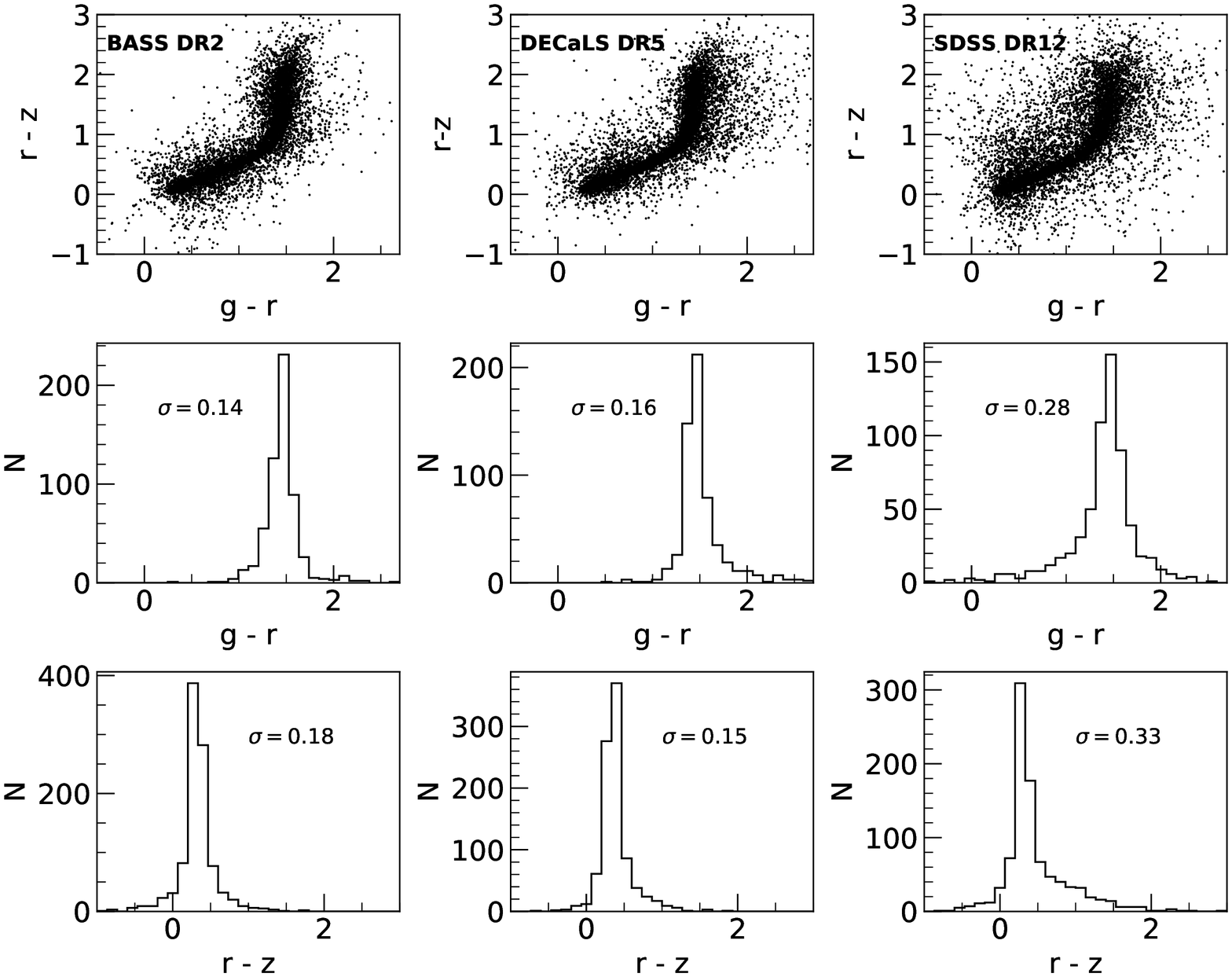}
\caption{Top row: color-color diagrams in $g - r$ vs. $r - z$. Middle row: $g - r$ color distribution at $r - z \sim 1.6$. Bottom row: $r - z$ color distribution at $g - r \sim 0.7$. From left to right, the panels display BASS DR2, DECaLS DR5, and SDSS DR12, respectively.}\label{fig-ccd}
\end{figure*}

The DECaLS has released its DR5 in 2017 June. Its typical depths with three exposures are 24.7, 23.6 and 22.8 mag for the $g$, $r$, and $z$ bands, respectively. The $g$-band and $r$-band depths are a little deeper than those of BASS DR2, while the $z$ band is somewhat shallower. There are some overlaps between DECaLS and BASS observations. These two surveys almost cover the whole SDSS footprint, but they are 1.5--2.5 mag deeper. We show some photometric comparisons of the three surveys with the data in the overlapping area. Figure \ref{fig-magcomp} shows the magnitude difference as function of magnitude for point sources, which are randomly selected based on the morphological type in DECaLS DR5. The DECaLS and SDSS data are both transformed to the BASS/MzLS photometric systems using different color terms. The photometric scatters of point sources brighter than 20 mag between DECaLS and BASS are about half of those between SDSS and BASS. We also compare the color-color diagrams of three surveys with the same dataset ($r < 23$ mag), as shown in the upper panels of Figure \ref{fig-ccd}. The color distributions of $g - r$ at $r - z \sim 1.6$ and  $r - z$ at $g - r \sim 0.7$ are also shown in the middle and bottom panels of Figure \ref{fig-ccd}. The standard deviations of the color distributions are annotated in each panel. From these plots, we can see that BASS DR2 is as good as DECaLS DR5, and it is obviously better than the SDSS data. 

\section{Data Access and Guidelines for DR2}  \label{sec-data}
\subsection{Data Access}
\begin{figure*}[!ht]
\centering
\includegraphics[width=0.8\textwidth]{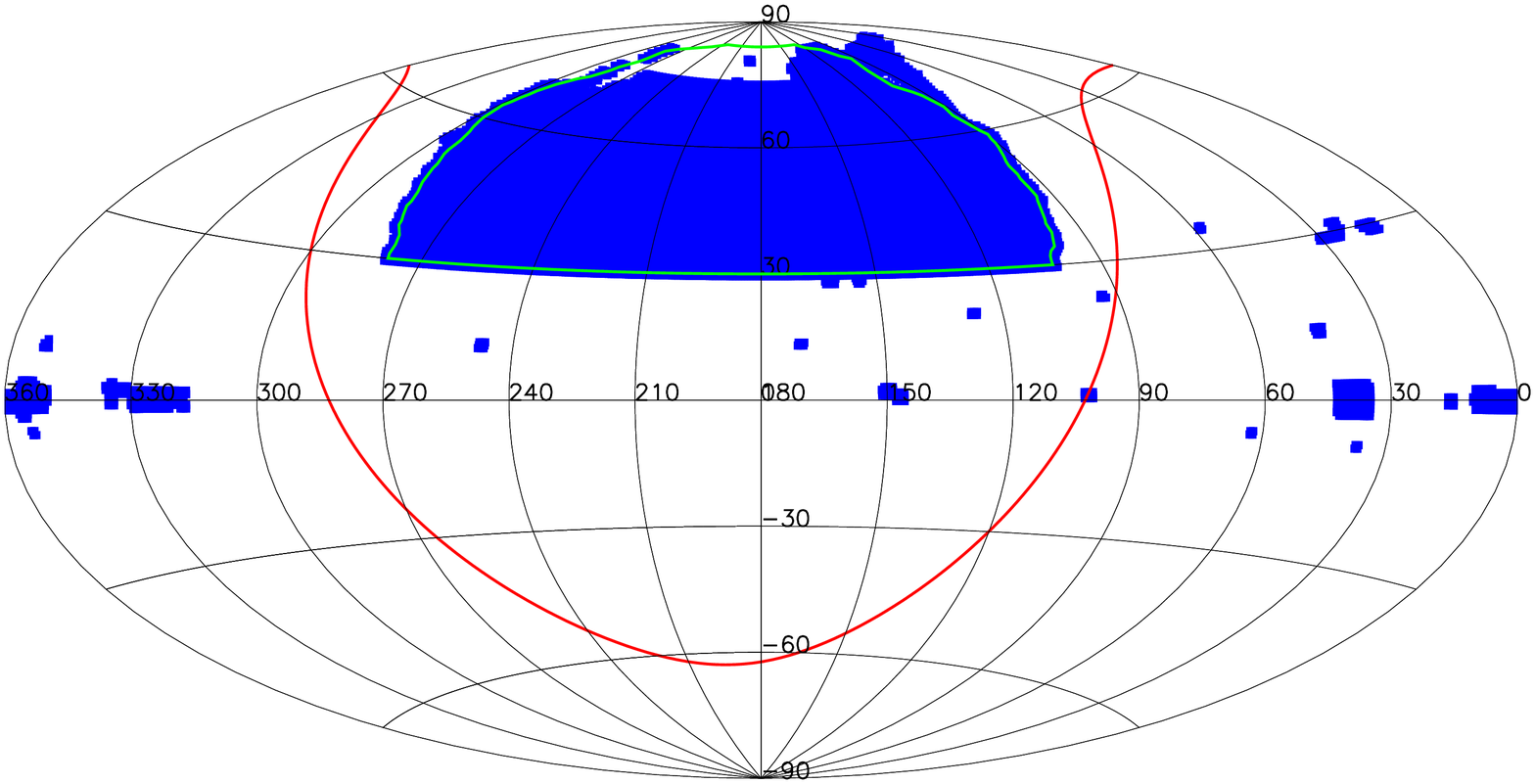}
\caption{Sky coverage of the DR2 co-added catalogs in Aitoff projection. The region outlined in green indicates the regular survey area. The red curve shows the Galactic plane. }\label{fig-catalog}
\end{figure*}

Information about the surveys, such as the telescopes, instruments, and observations, can be retrieved on the public data release website\footnote{\url{http://batc.bao.ac.cn/BASS/doku.php?id=datarelease:}}. The data cover the observations between 2015 January and 2017 July. There are 297,367 calibrated single-epoch CCD images. Those images with zero-point calibrations have corresponding catalogs. There are 13,419 blocks in total that have stacked images. For each block, there are three stacked images for the $g$, $r$, and $z$ bands and one co-added catalog. The summary files for single-epoch images and blocks can be found on the webpage\footnote{\url{http://batc.bao.ac.cn/BASS/doku.php?id=datarelease:dr2:home#data_access}}. Figure \ref{fig-catalog} shows the sky coverage of the DR2 catalogs. In addition to the regular survey area as shown in a green envelope, there are some test regions (e.g. COSMOS and Stripe 82) and scattered regions with observations shared with BASS by other observers.

Access to DR2 data is provided by the Chinese Virtual Observatory. Select data can be acquired from the following links.
\begin{enumerate}[1.]
   \item DR2 release webpage: \url{http://batc.bao.ac.cn/BASS/doku.php?id=datarelease:dr2:}
   \item To search for single-epoch image and catalog files: \url{http://explore.china-vo.org/data/bassdr2images/f}
   \item To search for stacked image and co-added catalog files: \url{http://explore.china-vo.org/data/bassdr2stack/f}
   \item To search for sources in co-added catalogs: \url{http://explore.china-vo.org/data/bassdr2coadd/f}
   \item To generate ``wget" scripts for downloading files: \url{http://batc.bao.ac.cn/BASS/doku.php?id=datarelease:dr2:dr2_wgetbulk:}
   \item Sky viewer for visualizing the data: \url{http://batc.bao.ac.cn/BASS/doku.php?id=datarelease:dr2:dr2skyviewer:}
\end{enumerate}
The sky viewer is based on Aladin HiPS \footnote{\url{http://aladin.u-strasbg.fr/hips/}}.

\subsection{A few guidelines for using the data}
\subsubsection{Images}
The calibrated single-epoch images have astrometric and photometric solutions stored in the FITS header. The flux units in the images are e/s. For each image, we have corresponding weight and flag images. The flag image tags the problematic pixels, the same as in DR1. We summerize useful parameters for single-epoch images in a FITS table \footnote{\url{http://batc.bao.ac.cn/BASS/lib/exe/fetch.php?media=datarelease:dr2:bassmzls-dr2-ccdinfo.fits}} and the user can find the column description on the wiki page\footnote{\url{http://batc.bao.ac.cn/BASS/doku.php?id=datarelease:dr2:dr2_ccdinfo:home}}. The single-epoch images with the column ``imq" $=$  1 are used for generating stacked images and co-added catalogs. There are several columns, as shown in Table \ref{tab-ccdinfo}, that denote astrometry qualities and zero-point. The ``astr\_ref" have two options: GAIA-DR1 and 2MASS. Usually, the astrometric accuracy obtained from using 2MASS as the reference catalog is worse than that using the Gaia catalog. The internal and external zero-points are combined as ``zp", which is used in our photometry. The magnitude can be calculated as $m = -2.5\mathrm{log}_{10}F + \mathrm{zp}$, where $F$ is the integrated flux of an object. Note that the single-epoch images have both astrometric and photometric residuals even though we can correct them in the catalogs. If the user wants higher accuracies for the astrometry and photometry from images, one should apply the residual maps to the coordinates and magnitudes\footnote{\url{http://batc.bao.ac.cn/online-data/BASS/dr2-residual-maps}}. The astrometric correction can be as much as 0\arcsec.04 for both BASS and MzLS. The photometric correction can be as much as 0.02 and 0.05 mag for BASS and MzLS, respectively.  
\begin{table*}[!ht]
\centering
\caption{Key Columns Describing Astrometry and Zero-point in the Summary File of CCD Images} \label{tab-ccdinfo} 
 \begin{tabular}{r|c|l}
\tableline\tableline
Column & Unit & Description  \\
 \tableline 
ra\_off	& 	arcsec	& R.A. offset relative to the reference catalog \\
dec\_off	& 	arcsec	& Decl. offset relative to the reference catalog \\
ra\_rms	&	arcsec	& R.A. rms relative to the reference catalog \\
dec\_rms	&	arcsec	& Decl. rms relative to the reference catalog \\
astr\_num	&	\nodata	& Number of stars used for statistics of astrometric parameters \\
astr\_ref	&	\nodata	& Reference catalog for astrometry \\ 
zp	        &	mag	        & Combined zero-point \\
zp\_rms	&	mag	        & Combined zero-point rms \\
zp\_num	&	\nodata    &  Number stars for combined zero-point \\
\tableline 
\end{tabular}
\end{table*}

The stacked images cannot be used for accurate photometric measurements because they are combined from many single-epoch images with different qualities. Each stacked image has a weight image. We adopt standard WCS parameters with a simple ``TAN" projection. The zero-points are fixed to 30. 

\subsubsection{Catalogs}
The co-added catalogs provide the deepest and most accurate magnitude measurements for objects with multiple exposures. A co-added catalog covers an area of about 0.68$\times$0.68 deg$^2$ with an 0\arcdeg.02 overlap with other neighboring blocks. For each source in the catalog, we provide the minimums, maximums, and averages of modified Julian date, seeing in FWHM, sky brightness, and zero-point from single-epoch observations, and provide the total exposure time and number of exposures in each band. The catalogs include Galactic reddening from the map of \citet{sch98}, some measurements from stacked images derived by SExtractor, such as shape parameters and star/galaxy classification (marked with a postfix ``\_Stack"), and ``Mag\_Auto" magnitudes for three bands. These measurements are only recorded and much less accurate. The other measurements in the catalogs are obtained by co-adding parameters from single-epoch images by our own photometric code. For each band, there are coordinates, shape parameters (e.g. half major/minor axis length, ellipticity, PA and Kron radius), all kinds of fluxes and magnitudes, and flags. For the magnitude error, we provide both normal parameter error and standard error. The standard error is related to the magnitude rms of multiple observations. It contains information of light variation, which can be useful for variable objects. Shape parameters are directly computed from the images. They are not intrinsic due to the seeing effect. The single-epoch catalogs contains the measurements for objects detected on stacked images. The object ID is unique in the co-added catalogs, while objects in single-epoch catalogs use the same ID. This means that the ID in single-epoch catalogs can be duplicated for objects with multiple observations. All fluxes in the catalogs are in nanomaggy, where magnitudes can be calculated as $-2.5\mathrm{log}_{10}F + 22.5$. We recommend the PSF magnitude for point sources and Kron magnitudes for extended sources. For very large galaxies with apparent diameters larger than about 1\arcmin, they might be fragmented due to foreground stars and their substructures, which might mean the photometry might not be reliable. %\textbf{A flag will be provided in DR3 to represent those sources that are contaminated by bright stars and large galaxies or nebulae.}

For star/galaxy separation, the column ``Type\_Stack" gives a median of class parameters from stacked images derived by SExtractor. We can also identify point sources from the reduced $\chi^2$ of PSF fitting. The reduced $\chi^2$ can be computed as the ratio of ``Chi2\_PSF" and ``DOF\_PSF" columns in the catalogs. If it is close to 1, the object is likely to be a point-like source. In addition, the ellipticity, color, and magnitude difference between PSF and Kron magnitudes can be also useful for classification. The user can even combine all these parameters to separate objects. We expect a classification to be added in the next data release. 

\section{Summary} \label{sec-sum} 
The BASS made its first data release be public in 2017 January. It only includes the BASS $g$-band and $r$-band observations taken before 2016 July.  This paper describes the details of our second data release, which include new datasets and updates to the data reduction. We summarize these updates as follows:

\begin{enumerate}[(a) ]
\item The DR2 includes the data taken as of 2017 July. This release includes the MzLS data. The BASS and MzLS have respectively completed about 72\% and 76\% of their observations over the footprint of over 5000 deg$^2$. 
\item BASS and MzLS data are reduced by the same pipeline, which includes some updates from DR1, such as bias correction of 2015 BASS data, identification of cosmic rays, crosstalk correction, and subtraction of pattern noise in MzLS images.
\item Gaia DR1 catalogs are utilized to derive astrometric solutions. The global astrometric error is about 0\arcsec.03. This is much better than DR1, where SDSS DR9 was used as the reference catalog. 
\item External photometric zero-points are calculated using point-source catalogs of PS1, the same as in DR1. To improve the accuracy of flux calibrations, internal zero-points are computed by comparing the magnitude differences of common objects in different exposures. 
\item Source detection is implemented in stacked images and photometry is made in single-epoch images. We provide circular aperture, Isophotal, Kron elliptical aperture, and PSF magnitudes. The median depths over the whole coverage are 24.05, 23.61, and 23.10 mag for the $g$, $r$, and $z$ bands, respectively. 
\item The DR2 data products include calibrated single-epoch images, single-epoch catalogs, stacked images, and co-added catalogs. They can be accessed through links in the BASS data release webpage (\url{http://batc.bao.ac.cn/BASS/doku.php?id=datarelease:dr2:}).
\end{enumerate}

The MzLS finished its observations in February of 2018 and the BASS will complete the observations in of 2019 January. The next data release (DR3) is expected to include all BASS and MzLS survey data. Furthermore, we will have some new features in DR3, such as astrometry by \textit{Gaia} DR2, a new source detecting algorithm, a star-galaxy separation, a flag to mark objects close to bright stars and large galaxies, new PSF modeling, and Galactic extinction from \textit{Planck} dust map.

\acknowledgments
We thank the anonymous referee for providing a rapid and thoughtful comments that improve our paper greatly. The BASS is a collaborative program between the National Astronomical Observatories of the Chinese Academy of Science and Steward Observatory of the University of Arizona. It is a key project of the Telescope Access Program (TAP), which has been funded by the National Astronomical Observatories of China, the Chinese Academy of Sciences (the Strategic Priority Research Program ``The Emergence of Cosmological Structures" Grant No. XDB09000000), and the Special Fund for Astronomy from the Ministry of Finance. The BASS is also supported by the External Cooperation Program of Chinese Academy of Sciences (grant No. 114A11KYSB20160057) and the Chinese National Natural Science Foundation (grant No. 11433005). The BASS data release is based on the Chinese Virtual Observatory (China-VO).

This work is also supported by the Young Researcher Grant of National Astronomical Observatories, Chinese, Academy of Sciences, the Chinese National Natural Science Foundation (grants No. 11673027, 11733007, 11603034, 11503051, U1531115, and U1731243), and the National Basic Research Program of China (973 Program; Grants No. 2015CB857004, 2014CB845704, 2014CB845702, and 2013CB834902). L.J. acknowledges support from the National Key R\&D Program of China (2016YFA0400703) and from the Chinese National Science Foundation (grant No. 11533001). A.D.M. and J.R.F. were supported by the Director, Office of Science, Office of High Energy Physics of the U.S. Department of Energy under Contract No. DE-AC02-05CH1123.

The Pan-STARRS1 Surveys (PS1) and the PS1 public science archive have been made possible through contributions by the Institute for Astronomy, the University of Hawaii, the Pan-STARRS Project Office, the Max-Planck Society and its participating institutes, the Max Planck Institute for Astronomy, Heidelberg and the Max Planck Institute for Extraterrestrial Physics, Garching, The Johns Hopkins University, Durham University, the University of Edinburgh, the Queen's University Belfast, the Harvard-Smithsonian Center for Astrophysics, the Las Cumbres Observatory Global Telescope Network Incorporated, the National Central University of Taiwan, the Space Telescope Science Institute, the National Aeronautics and Space Administration under grant No. NNX08AR22G issued through the Planetary Science Division of the NASA Science Mission Directorate, the National Science Foundation Grant No. AST-1238877, the University of Maryland, Eotvos Lorand University (ELTE), the Los Alamos National Laboratory, and the Gordon and Betty Moore Foundation.

This work has made use of data from the European Space Agency (ESA) mission {\it Gaia} (\url{https://www.cosmos.esa.int/gaia}), processed by the {\it Gaia} Data Processing and Analysis Consortium (DPAC, \url{https://www.cosmos.esa.int/web/gaia/dpac/consortium}). Funding for the DPAC has been provided by national institutions, in particular the institutions participating in the {\it Gaia} Multilateral Agreement.

This publication makes use of data products from the Two Micron All Sky Survey, which is a joint project of the University of Massachusetts and the Infrared Processing and Analysis Center/California Institute of Technology, funded by the National Aeronautics and Space Administration and the National Science Foundation.

SDSS-III is managed by the Astrophysical Research Consortium for the Participating Institutions of the SDSS-III Collaboration including the University of Arizona, the Brazilian Participation Group, Brookhaven National Laboratory, Carnegie Mellon University, University of Florida, the French Participation Group, the German Participation Group, Harvard University, the Instituto de Astrofisica de Canarias, the Michigan State/Notre Dame/JINA Participation Group, Johns Hopkins University, Lawrence Berkeley National Laboratory, Max Planck Institute for Astrophysics, Max Planck Institute for Extraterrestrial Physics, New Mexico State University, New York University, Ohio State University, Pennsylvania State University, University of Portsmouth, Princeton University, the Spanish Participation Group, University of Tokyo, University of Utah, Vanderbilt University, University of Virginia, University of Washington, and Yale University.

\end{document}